\documentclass[12pt]{iopart}
\usepackage[dvips]{graphicx,epsfig}

\begin{document}
\article[Strange hadron ratios from quark coalescence
at RHIC and LHC ]{}
{Strange hadron ratios from quark coalescence \\ at RHIC 
and LHC energies}
\date{30 October 2007}
\author{ P. L\'evai$^a$}
\address{$^a$ RMKI Research Institute for Particle and Nuclear Physics, \\
P.O. Box 114, Budapest, 1525, Hungary}

\begin{abstract}
Quark coalescence models have been applied successfully to reproduce measured
hadron production data in relativistic heavy ion collisions at SPS and
RHIC energies, which finding strongly supports the formation of deconfined
quark matter in these collisions. 
The investigation of meson and baryon production is an ideal tool to
understand dynamical details of hadronization,
especially strange hadron numbers and ratios.
We display latest results on the production of  strange
particles in quark coalescence processes in heavy ion collisions
at different collider energies.
\end{abstract}

After many years of investigating hadron-hadron and heavy ion collisions the 
study of hadron production remained an important research
field.  The lack of perfect knowledge of the microscopical mechanisms
led to the application of many different models, very often from
completely opposite directions. Statistical models
are based on the introduced statistical weights for produced
hadrons~\cite{Fermi50,Pomer51,Haged65}. This idea has been developed
further and thermal models appeared with the introduction of the
temperature parameter and thermal weights for hadrons
(see e.g. Ref.~\cite{Haged65}).
Thermal models became very successful reproducing soft particle
production in different high energy particle
collisions~\cite{Becatt}, especially in heavy ion collisions
(see e.g. Refs.~\cite{Koch83,SHARE,BM04,THERMUS}). 
On the other hand, these models assume the formation of a sort of
thermal and chemical equilibrium in the hadronic phase and 
determine thermodynamical variables for the 
produced hadron phase. The deconfined period of the time evolution 
dominated by quarks and gluons remains hidden: full equilibration generally
washes out and destroys large amount of information about the early deconfined phase.
The success of statistical models implies the existence of such
an information loss during hadronization, at least for certain properties.
It is a basic question which properties can survive the hadronization
and behave as messengers from the early (quark dominated) stages.

Since our main goal is to create quark matter in heavy ion collisions
and determine its properties, we can not be satisfied even with the
most perfect hadronic statistical model. 
We need to survey the messengers of the early phases and investigate
hadron production by models based on quark degrees of freedom. 
Hadronization models with direct quark degrees of freedom have been 
constructed from the beginning of the heavy ion programme at CERN SPS.
In this stream quark coalescence has been proposed many years ago
to describe quark matter hadronization~\cite{hwa77,alcor,alcor2,bialas02}. 
Massive constituent quarks has been considered in the 
deconfined phase, which quarks and anti\-quarks
are ready to hadronize through "coalescence", a clustering process 
driven by an attractive 
force between the properly coloured quark degrees of freedom. 
The analysis of lattice QCD data has supported the 
presence of such massive excitations in the
quark-matter phase close to the quark-hadron
phase transition~\cite{PRC98LH}. The attractive force generated by
gluons (which are considered via this interaction and ignored
as independent degrees of freedom), and it is modeled
by non-relativistic colour potential between the 
massive quarks and antiquarks.
Mesons are produced by quark-antiquark coalescence. Baryons
are produced in two steps: at first $\overline{3}$ diquarks appear through
quark-quark coalescence, which is followed by a diquark-quark
coalescence into a colourless baryon. 
Thus we have microscopical steps to describe hadronization and
proper physics to fulfill conservation laws.

Particle yields, ratios and spectra measured in heavy ion collisions
have been reproduced successfully in the ALCOR~\cite{ALCORdat1,ALCORdat2}
and MICOR~\cite{MICOR00,MICORSQM98} coalescence models.
However,  thermal models were similarly successful in the low-$p_T$ region 
and more widely used because of their simplicity.
At RHIC energy intense data collection has been performed in the
intermediate-$p_T$ region ($3 < p_T < 8$ GeV/c) and the
measured anomalous proton/pion ratio could have been explained by
quark coalescence and recombination models~\cite{hwa,greco,fries}. 
Since thermal models could not be applied for these data,
the production of relatively rear particles
(e.g. intermediate-$p_T$ hadrons, heavy flavours) became the main focus
of non-equilibrium models.
The interest in coalescence hadronization mechanism has been increased
and more applications appeared. 
One subfield is the strange hadron production: quark chemistry
based on strange quark coalescence
and annihilation became very successful to describe SPS and RHIC data.
Furthermore, quark flow production also moved into the
focus of interest. The recognition
of valence quark number scaling in asymmetric flow ($v_2$)
strongly supports quark matter formation 
and quark coalescence at RHIC and SPS energies~\cite{molnard}. 

It is natural to ask, if quark coalescence is working properly for heavier
flavour and at higher-$p_T$, then what about the soft region, namely
bulk hadron numbers, and ratios. During last decade we continously
summarized the results of quark coalescence calculations and the successful
reproduction of measured data, in parallel giving predictions where it was
possible. In this talk we gave a short summary again, analyzing latest
SPS and RHIC data and predict the expected particle yields at LHC
energy. We would like to emphasize that quark coalescence models
are capable to describe large amount of experimental data, in the meantime
supporting the formation of deconfined quark matter in heavy ion
collisions at SPS and RHIC energies.

Another question is connected to the validity of non-relativistic 
description in a highly relativistic strongly interacting particle 
ensamble~\cite{Hamar07}. 
Although all particles, including quarks in the early stage, are moving
with a velocity close to the speed of light, but in a comoving system they can 
coalesce if only their relative velocity is small. This self-regularization
gives the opportunity to describe the effective quark binding steps in
a non-relativistic frame with coalescence. 

\newpage
On the other hand, another description exists,
which is very similar in many ways: the light-front wave function (LFWF)
formalism~\cite{Brodsky} is based on constituent quarks moving along the
light-front and quantum mechanics is recovered in the comoving system.
It is interesting to note, that the LFWF formalism and the investigation
of quantum-chromodynamics on light-front displays a connection to the
AdS/CFT correspondence.  Bound states of relativistic massless quarks can be 
described in this formalism, and an effective Schr\"odinger equation
can be derived, where the effective potential is dictated by
conformal symmetry~\cite{BrodskyADS}. 
This correspondence is beyond the scope of our recent study and 
here we will stay at a phenomenological level including  
Schr\"odinger picture and quantum mechanics.

In quark coalescence models~\cite{alcor,MICOR00} a
premeson $h$ consists of quark $q_1$ and antiquark $q_2$, 
and production is proportional to densities of constituents,
$n_1$ and $n_2$:
\begin{eqnarray}
 \partial _\mu (n_h u^\mu)  =
 \langle \sigma^h_{12} v_{12} \rangle \ n_1 \, n_2 \ .
\end{eqnarray}
In an isotropic plasma state the rate, 
$\langle \sigma^h_{12} v_{12} \rangle$, is calculated
as a momentum average:
\begin{eqnarray}
\langle \sigma^h_{12} v_{12} \rangle &=& \frac{\int d^3 \vec{p}_1 d^3\vec{p}_2
\ f_q(m_1, \vec{p}_1) f_q(m_2, \vec{p}_2)  \sigma v_{12}}
{\int d^3 \vec{p}_1 d^3 \vec{p}_2 \ f_q(m_1,\vec{p}_1)f_q(m_2,\vec{p}_2)}
\end{eqnarray}
The quark coalescence cross section is determined from
quantum mechanics, assuming a rearrangement ("pick-up") reaction~\cite{Schiff}.
In the ALCOR model~\cite{alcor} quark plain waves coalesce into a bound two-body
system, described by a hydrogen-like wave function. Coalescence process
is driven by a Coulomb-potential depending on relative distance, $r$:
\begin{equation}
V(r)=\frac{{\alpha}{\langle \lambda_i\lambda_j \rangle} }{r}
\end{equation}
The colour factor 
${\langle \lambda_i\lambda_j \rangle}$ is determined by the colour
combination of the interacting particles. This factor is $-4/3$
for quark-antiquark coalescence into a color singlet pre-meson,
and $-2/3$ for quark-quark coalescence into an antitriplet (${\overline 3}$)
diquark state.

In a parallel study~\cite{Hamar07}
we have investigated the robustness of the quantum mechanical description,
applying different wave-function setups and using Yukawa potential 
with wide region for the screening mass. We demonstrated that the final
results for hadron ratios are very close to each other, although we
have very much different intermediate values for coalescence rates
and quark ratios obtained from a fitting procedure.
Thus we will use here the ALCOR model's wave function setup, 
and we display the obtained results. Other wave function setup may
yield slightly different hadron ratios, which will be surveyed
in the near future and reported elsewhere.

Here in the ALCOR model there are four input parameters.
Entropy production is followed by the number
of newly produced light quark-antiquark pairs 
($N_{u{\overline u}} = N_{d{\overline d}} $). Strangeness production
is controlled by the number of newly produced strange 
quark-antiquark pairs
indicated by the ratio $f_s= N_{s{\overline s}}/(N_{u{\overline u}} +
N_{d{\overline d}})$.
Baryon to meson ratio is followed by the $\alpha_s$ effective coupling
constant, because meson production is proportional to $\alpha_s$, but
two-steps baryon production depends on $\alpha_s^2$.
Particle to antiparticle ratios are controlled by the stopping 
of incoming colliding nucleons into
the mid-rapidity region.

\begin{table}
\caption{
ALCOR results at SPS energies, 
$E_{\rm beam}=$ 20, 30, 40, 80, 158 GeV/n.
}
\begin{indented}
\item[]\begin{tabular}{@{}l|lllll}
\br
dN/dy at y=0       &  20 GeV/n & 30 GeV/n  & 40 GeV/n  & 80 GeV/n & 158 GeV/n   \\
\br
Input data          &&&&& \\
$\pi^-$             & 85 $\pm$ 5 & 96 $\pm$ 5 & 110 $\pm$ 5 & 145 $\pm$ 5 & 182 $\pm$ 5 \\
$K^-$               &5.6 $\pm$ 0.2& 7.8 $\pm$ 0.2& 8 $\pm$ 0.5& 12 $\pm$ 0.5& 17.5 $\pm$ 0.5 \\
$K^+/K^-$           &3.0 $\pm$ 0.2& 2.6$\pm$0.2&2.55$\pm$0.2& 2.1$\pm$ 0.2& 1.7$\pm$ 0.2 \\
${\overline \Xi}^+$ &--- &0.05 $\pm$ 0.02 & 0.07 $\pm$ 0.02 & 0.2 $\pm$ 0.05 &0.34 $\pm$ 0.04 \\
\br
ALCOR param.        &&&&& \\
New $u{\overline u}$  & 45 & 50 & 62 & 88 & 123    \\
$f_s$ strangeness          &\ 0.40 &\ 0.35&\ 0.30 &\ 0.28 &\ 0.24 \\
Stopping     &20 \%& \ 20 \%& \ 20 \%& \ 20 \%& \ 15 \% \\
$\alpha_s$ coupling   &\ 0.80 &\ 0.80& \ 0.80& \ 0.80&\ 0.72 \\
New \# ${Q\overline Q}$     & 252 & 270  & 322  & 450 & 610   \\
\br
ALCOR results    &&&&& \\
$K^+/\pi^+$   &0.213 & 0.192 & 0.173 & 0.171 & 0.158  \\
${\overline \Lambda}^0/{\overline p}^-$&
              \ 1.36 & \ 1.16& \ 1.0 & \ 0.93 & 0.80  \\
$\Phi/K^-$    &0.42 & 0.385& 0.34  & 0.32  & 0.28  \\
$(\Omega^-+ \Omega^+)/\pi^-$  &
              0.00268  & 0.00217 & 0.00189 & 0.00233 & 0.00244  \\
\br
\end{tabular}
\end{indented}
\end{table}

In Table 1 the latest results from quark coalescence calculations
are summarized at different  CERN SPS energies in the mid-rapidity
regions of $Pb+Pb$ collisions~\cite{CERN1,CERN2}. 
The four model parameters are determined from four experimental
data, especially from $\pi^-$ and $K^-$ numbers indicating entropy
and strangeness production, from $K^+/K^-$ ratio indicating
particle/antiparticle ratio, and from the absolute number of
${\overline \Xi}^+$ particle connected to the effective coupling constant. 
(At 20 GeV/n no data were available,
in this case $\Xi^-$ has been used.)
Table 1 displays that entropy production is
increasing, but relative strangeness abundance ($f_s$) is continously
decreasing with increasing collision energy. 
One can see, the ratio $K^+/\pi^+$
is decreasing, as well as the ${\overline \Lambda}^0/{\overline p}^-$
and $\Phi/K^-$ ratios,  as we expect from the decreasing of $f_s$.
The $K^-/\pi^-$ ratio is slightly increasing, but this is 
an exception.  The energy dependence of
$\langle \Omega \rangle / \pi^-$ has a special structure, which
should be investigated in details.
The stopping is
close to be constant, it starts to decrease at highest SPS energy,
as well as the effective coupling constant.

We repeat our analysis at RHIC energies. Table 2 displays recent ALCOR results
at $\sqrt{s}= 200$ AGeV in $Au+Au$ collisions. We can see that
entropy production is increasing further, but strangeness production
is saturated at a certain value ($f_s=0.22$), which number is valid
at $\sqrt{s}= 130$ AGeV, also. The coupling constant is also
saturated around $\alpha_s=0.55$, founded at lower RHIC energy.
The stopping is decreasing, as it is indicated by close to unity
antibaryon to baryon ratios.
The quark coalescence model can reproduce bulk particle ratios in most 
of the cases, except the $\Phi/K^-$. We investigate recently
if different wave-function setups could improve this agreement~\cite{Hamar07},
keeping other ratios under good control. Another way to improve
the quark coalescence results is to investigate the role of
higher baryon and meson resonance production channels.

\begin{table}
\caption{
ALCOR results at RHIC energy ($\sqrt{s}=200$ AGeV)
 and predictions for LHC energy ($\sqrt{s}=5500$ AGeV).
}
\begin{indented}
\item[]\begin{tabular}{@{}l|ll|ll}
\br
Particles       &  Data  & ALCOR  & ALCOR  & ALCOR   \\
(dN/dy at y=0)  &  RHIC  & RHIC   & LHC-I  & LHC-II  \\
\br
New $u{\overline u}$  && 286   & 500   & 750    \\
$f_s$ strangeness          && \ 0.22& \ 0.22& \ 0.22 \\
$\alpha_s$ coupling   && \ 0.55& \ 0.55& \ 0.55 \\
Stopping            && \ 3 \%& \ 1 \%& \ 1 \% \\
Total \# ${Q\overline Q}$   && 1396  & 2440  & 3660   \\
\br
$h^{\pm}$     & 780 $\pm$ 40  & 780   & 1252     & 1830  \\
$\pi^-$       & 327 $\pm$ 32  & 322   & \ 500    & \ 724  \\
$K^+$         & 51.3 $\pm$ 7.7& \ 48  & \ \ 70   & \ \ 99  \\
$p^+$         &               & \ 19  & \ \ 37   & \ \ 62  \\
$\Xi^-$       &2.16 $\pm$ 0.09& \ 2.59& \ \ 6.42 & \ \ 10.7 \\
$K^+/\pi^+$   &0.16 $\pm$ 0.02& \ 0.15& \ \ 0.14 & \ \ 0.14  \\
$\Xi^-/\pi^-$ &0.007$\pm$0.001& \ 0.008&\ \ 0.013& \ \ 0.015  \\
$\rho^0/\pi^0$&0.20 $\pm$ 0.04& \ 0.22& \ \ 0.21 & \ \ 0.20  \\
$\Phi/K^-$    &0.15 $\pm$ 0.03& \ 0.26& \ \ 0.25 & \ \ 0.25  \\
\br
\end{tabular}
\end{indented}
\end{table}

Combining quark coalescence results at SPS and RHIC energies, we can
predict different particle yields at LHC energies.
We can expect from Table 1 and 2, that strangeness production and 
effective coupling constant will be very similar at RHIC and LHC energies,
on the other hand baryon number stopping into the mid-rapidity
region will drop to a minimal value (e.g. 1 \%).
The only question is connected to the value of entropy production, especially
to the new light quark-antiquark pair production. Figure 1. displays the
obtained values at SPS and RHIC energies and indicates
a linear increase in entropy production with increasing $\ln \sqrt{s}$.
Applying a linear extrapolation we obtain the estimated value 
$N_{u\overline u}/dy=500$ for LHC energies in $Pb+Pb$ collisions
at mid-rapidity. Table 2 displays the ALCOR results for this 
light quark pair production, keeping the other parameter values,
 as we discussed above.
In order
to test the sensitivity of some ratios to the amount of produced
entropy, we give, in the last column of Table 2, results of our
model assuming a 50 \% higher entropy.
Many strange particle ratios are
insensitive on the higher entropy production, thus other data are
necessary to investigate entropy production at LHC energies,
e.g. absolute particle numbers. 

\begin{figure}
\begin{center}
\includegraphics[width=7.7cm]{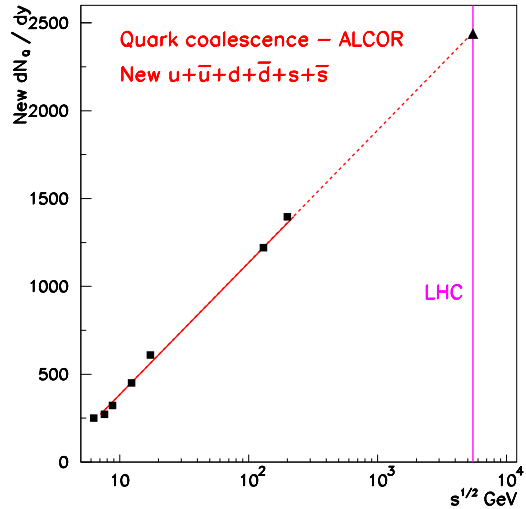}
\end{center}
\caption{\protect\label{newq} The energy dependence
of the newly produced quark and antiquarks in the mid-rapidity 
region. The displayed values are obtained from the
ALCOR model fit to the experimental data.
The value at LHC energy is obtained from a linear extrapolation.
} 
\end{figure}

This work has been supported in part by 
the Hungarian OTKA under grants No. NK062044 and IN71374.

\end{document}